\documentstyle[preprint,epsf]{jpsj}

\newcommand{\r}{{\mib r}}
\newcommand{\p}{{\mib p}}
\newcommand{\q}{{\mib q}}

\newcommand{\en}{\varepsilon}
\newcommand{\om}{\omega}

\def\runtitle{Instructions for the Preparation of Manuscript}
\def\runauthor{Hiroto {\sc Adachi} and Ryusuke {\sc Ikeda}}

\title
{
Fluctuation Conductivity in Unconventional Superconductors near Critical 
Disorder\\
}

\author
{Hiroto
 {\sc Adachi}  and  Ryusuke {\sc Ikeda}}

\inst
{
Department of Physics, Kyoto University, Kyoto 606-8502\\
}

\recdate
{
May 24, 2001
}

\abst
{
The fluctuation conductivity $\sigma_{\rm s}$ in bulk superconductors 
with non $s$-wave pairing and with nonmagnetic disorder of 
strength $D$ is studied at low $T$ and within the Gaussian 
approximation. 
It is shown by assuming a quasi two-dimensional (2D) 
electronic state that, only if the gap function 
$d_\mu({\p})$ is, as in a 2D $p$-wave pairing state, linear in the 
in-plane (relative) momentum ${\p}_\perp$, the in-plane fluctuation 
conductivity on the line $D=D_c$ is weakly divergent in low $T$ limit. 
The present result may be useful in clarifying the true gap 
function of spin-triplet ${\rm Sr_2RuO_4}$ through resistivity measurements. 
}

\kword
{
 Sr$_2$RuO$_4$,fluctuation conductivity,
unconventional superconductivity, dirty superconductor
}

\begin{document}
\sloppy
\maketitle


Conventionally, the pairing state of a superconducting (SC) 
phase is identified by examining microscopic properties 
deep in the SC phase or by directly finding a pairing mechanism 
consistent with the electronic state of the material.
In contrast, it is usually difficult to obtain information on the pairing
state through physical properties near a SC transition such as 
the fluctuation effects. 
For example, the fluctuation conductivity $\sigma_{\rm s}$
accompanying a {\it thermal} transition depends only on the 
dimensionality\cite{Tink} of fluctuation except for a constant 
numerical prefactor of its expression, which is not\cite{ROT,rf:Emery75} 
easy to examine experimentally with high precision. 
But how about near a SC transition at {\it zero} temperature ($T=0$) ? 
From this point of view, we examine the nature of $\sigma_{\rm s}$ in 
unconventional (non $s$-wave) superconductors 
near a quantum critical point, $(D, T) = (D_{\rm c}, 0)$, which is the end 
point of disorder-driven transition line $T_{\rm c}(D)$, where $D$ is the strength 
of microscopic (nonmagnetic) disorder. 
Particular attention is paid to $\sigma_{\rm s}$ on the specific line $D=D_{\rm c}$ 
in the $D-T$ phase diagram. 
We find in the Gaussian approximation that, only in the case of pairing 
of $p$-wave type,\cite{com} $\sigma_{\rm s}$ on the $D=D_c$ line is weakly 
divergent in low $T$ limit. 

This work was motivated by the unresolved issue of the pairing 
symmetry of ${\rm Sr_2RuO_4}$\cite{rf:Maeno94}. 
In spite of various proposals, the 
pairing state of this material has not been determined thus far. 
A key issue is to determine where on the Fermi surface 
the nodes of SC energy gap are present.
Having ${\rm Sr_2RuO_4}$ in mind, we assume that the electronic states are
of quasi 2D character, while the dimensionality of low energy modes such as 
the superconducting fluctuation is three-dimensional (3D). 
Further, all calculations will be carried out within the weak coupling 
approximation, and the effects of disorder on the fluctuation, 
as well as on the normal properties, 
will be treated within the Born approximation 
since the nature of fluctuation is determined by the electronic details 
in the normal phase. 
A key fact in the present analysis is that,
when the gap function is of $p$-wave type (according to the classification 
defined in ref. 4), a part of the lowest order gradient terms in the 
Gaussian $\rm Ginzburg-Landau$ (GL) action is proportional to a 
$\ln(1/T\tau)$-factor arising from the diffusion propagator induced by 
the disorder, where $\tau$ is the lifetime of quasiparticles, 
while for gap symmetries with higher harmonics (i.e., $d$-wave, 
$f$-wave types\cite{com} or others) the corresponding $\ln(1/T\tau)$-factor 
appears only in higher gradient terms that are unnecessary for obtaining 
the leading term of $\sigma_{\rm s}$.

As a microscopic model, we start with the BCS Hamiltonian of a layered
material with a random potential term
\begin{eqnarray}\label{eq:hamiltonian}
	{\cal H}  
&=&
	\sum_\sigma \sum_{\p} \xi_{\mibs p} \, 
	a^\dagger_\sigma({\p}) a_\sigma({\p}) 
+
	\sum_{l}\int {\rm d}^2 \r_{\perp}\,
	u_l (\r_{\perp})a_{\sigma}^ \dagger(\r_{\perp}, l)
	a_{\sigma}(\r_{\perp}, l) \nonumber \\
&&-
	 \frac{1}{2}
	\sum_{\sigma,\sigma'}  \sum_{{\mibs p},{\mibs p}'{\mibs q}}
	V_{\rm BCS}(\p, \, \p')
	a_{\sigma}^\dagger(\p_+)
	a_{\sigma'}^\dagger(-\p_-)
	a_{\sigma'}(-\p'_-)
	a_{\sigma}(\p'_+), 
\end{eqnarray}
where $a_{\sigma} (\r_{\perp},l)$ is the annihilation operator of 
electron with spin $\sigma$ and at the in-plane coordinate 
$\r_{\perp}$ on the $l$-th plane, the attractive interaction 
$V_{\rm BCS}(\p, \, \p')$ is assumed to be, as usual, separable 
like $|g| w(\p) w\*(\p')$, and $\p_{\pm}$ implies $\p \pm \q/2$.
The random potential $u_l (\r_{\perp})$ obeys the
Gaussian ensemble: $\overline{u_l (\r_{\perp})}=0;
\overline{u_l (\r_{\perp})u_{l'} (\r_{\perp}')}=(2\pi N_2(0)\tau)^{-1}
\delta(\r_{\perp} - \r_{\perp}') \delta_{l,l'}$, 
where the overbar denotes the random average, and $N_2(0)$ is the 2D density 
of states at the Fermi surface. 
The impurity scattering was assumed at this stage to be isotropic. 
We have verified that an inclusion of $p$-wave scattering 
rate\cite{AGD} $\tau_1^{-1}$ does not affect the main features in 
eq. (\ref{eq:vertex}) becoming relevant to our derivation of 
$\sigma_{\rm s}(D=D_{\rm c})$. 

The random-averaged Green's 
function is, as usual, given by 
\begin{equation}\label{eq:green}
	{\cal G}(\p,\en_n)
=
	\frac{1}{{\rm i} \en_n (1 + (2 \tau |\en_n|)^{-1}) -\xi_\p}, 
\end{equation}
with a fermionic Matsubara frequency $\en_n$. 
The dispersion in eq.(\ref{eq:green}) takes 
the quasi 2D form
\begin{equation} \label{eq:dispersion}
	\xi_{\mibs p}
=
	\varepsilon_{{\mibs p}_\perp} - E_{\rm F} + t\cos(p_z s), 
\end{equation}
where $s$ is the interlayer distance and $\p=(\p_\perp, p_z)$.
Throughout this paper we take units $\hbar = k_{\rm B} = 1$.
Also we neglect the localization effect on noninteracting
electrons and treat the disorder effect within the Born approximation.

To illustrate the points essential to our results, let us first explain
the analysis of a case of $d$-wave type defined by 
$V_{\rm BCS}=|g|w_d(\p) w^*_d(\p')$ 
with 
$w_d(\p) = (p_x^2-p_y^2)h_d(p_zs)$. 
Here, $h_d(p_zs)$ is assumed to satisfy
$\langle h_d \rangle = s \int_{p_z} h_d(p_zs) > 0$ 
so that $h_d$ may be $p_z$-independent,
reflecting the quasi 2D dispersion (\ref{eq:dispersion}). 

\begin{figure}[t]
\begin{center}
 \leavevmode
 \epsfysize=2cm
 \epsfbox{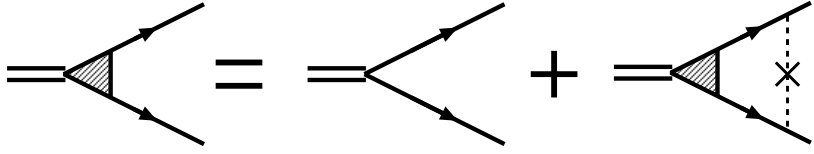}
\end{center}
 \caption{
Feynman diagrams defining the renormalized vertex 
$W_{\q}({\p})$ (the hatched region), 
where the double line, the solid line with arrow, 
and the dashed line denote, respectively, the pair-field, 
the electron propagator, and the impurity line carrying $(2 \pi \tau N_2(0))^{-1}$. 
The first term of r.h.s. implies the bare vertex $w_d({\p})$. 
}
\label{fig:fig1}
\end{figure}

Following the standard procedure\cite{KAE} 
to decoupe the BCS interaction in terms of 
the pair-field $\Delta_\mu$ which is related to the $d$-vector as 
$d_{\mu}(\p) = w_d(\p) \Delta_{\mu}$, where $\mu$ is the spin index, 
the Gaussian GL action takes the form 

\begin{equation} \label{eq:actioin}
	S
=
	\sum_{\mu} \sum_{q,\om_{\nu}} {\cal K}^{-1}({\q}, \om_\nu)
|\Delta_{\mu}(\q,\om_{\nu})|^2, 
\end{equation}
where ${\cal K}^{-1} = (|g|N_2(0))^{-1} - \pi({\q},\om_{\nu})$, and 

\begin{equation} \label{eq:pi}
	\pi({\q}, \om_\nu) 
=
        T  \sum_{\en_n} \int {\rm d} \epsilon_{\mibs p_\perp} 
	\int \frac{{\rm d}(p_zs)}{2\pi} 
        w^*_d(\p) W_{\mibs q}(\p) {\cal G}(\p_+, \en_{n+}) 
	{\cal G}(-\p_-, -\en_{n-}), 
\end{equation}
where the renormalized vertex $W_{\q}({\p})$ is defined 
in terms of the bare vertex $w_d({\p})$ as indicated by \cite{AGD} Fig. 1. 
In general, features of crystal structure appear in the dispersion 
$\varepsilon_{\mibs p_\perp}$ which may make the derivations of eq.(\ref{eq:pi}) and 
$W_{\q}({\p})$ complicated. If an isotropic Fermi surface is assumed, 
for simplicity, $W_{\q}({\p})$ is given by 

\begin{equation} \label{eq:vertex}
                W_{\q}(\p)
=
	w_d(\p) 
	- \frac{p_{\rm F}^2 l^2}{4}
	\langle  h_d \rangle (q_x^2 - q_y^2) \, \Gamma_q(2\en_n+\om_\nu), 
\end{equation} 
where 
$\Gamma_\q(2\en_n+\om_\nu)
=(\tau|2\en_n+\om_\nu|+l^2{q}_\perp^2/2 +2t^2\tau^2{\rm sin}^2(q_zs/2))^{-1}$ 
is the diffusion propagator, and $\om_{\nu}$ is a bosonic Matsubara frequency. 
Since the ${\q}$-dependence in the second term [i.e., $q_x^2-q_y^2$ 
in eq. (\ref{eq:vertex}) ] is essentially determined by 
the ${\p}$-dependence of $w_d$, the effect of crystal structure is, except 
for its role leading to the form of $w_d({\p})$, unimportant in 
deriving the renormalized part corresponding to the second term of eq.(\ref{eq:vertex}). 
An effect of crystal stucture on the GL gradient terms unaccompanied by the 
diffusion propagator 
will be commented on later together with an effect of the GL quartic term 
neglected here. 

It is easily seen that the last term of eq. (\ref{eq:vertex}) leads to higher
gradient terms like $\sim (q_x^2-q_y^2)^2$ in GL action. 
A $q_z$-dependence neglected above, which may appear in 
$\langle h_d \rangle$ of $W_{\q}$, is reflected only 
in a still higher gradient term like $(q_x^2-q_y^2)^2 q_z^2$. 
Since such higher gradient terms are irrelevant to the leading terms of interest 
in fluctuation conductivities, we can omit the last term of eq.(\ref{eq:vertex}) 
at this stage. 
Then, 
${\cal K}(\q, \om_\nu)$ simply becomes 

\begin{equation} \label{eq:propagator}
	{\cal K}^{-1}(\q, \om_{\nu})
= 	\mu_0 + \gamma_1 |\om_{\nu}| + \xi_{0}^2 q^2_\perp
	+4\left( \frac{\xi_{c0}}{s}\right)^2 \sin^2(q_z s/2). 
\end{equation}

Here $\mu_0= 2\pi^2 (T\tau)^2/3 + \left( D - D_{\rm c} \right)/ D_{\rm c}$
with $D/D_{\rm c}=1.78/(\pi T_{\rm c0} \tau)$, 
$\gamma_1 \simeq \tau$, $\xi_0^2=l^2/2$ with $l=v_{\rm F} \tau$, 
$\xi_{\rm c0}^2 = (t\tau s)^2/2$, 
and $T_{\rm c0}$ is $T_{\rm c}$ 
in clean limit $(D=0)$. 
Note that the fact $\int_{\mibs p_\perp}w_d(\p) = 0$ is essential for 
obtaining the $T_{\rm c}$-reduction induced by disorder. Hereafter, we only take 
the dissipative frequency term\cite{rf:Ikeda96} into account. 
Inclusion of other dynamical terms is found not to change our main 
conclusions. 
The above-mentioned feature that the contribution accompanied by 
the diffusion propagator is reflected 
only in higher gradient terms of GL action is common to all pairing states 
with harmonics higher (angular momentum) than those of p-wave type\cite{com}, 
and hence, the results from eq. (\ref{eq:sigma_s}), below, 
for the case of $d$-wave type are also valid for 
other states with higher harmonics including states 
of $f$-wave type. 

Next, let us calculate the fluctuation conductivity $\sigma_{\rm s\perp}$ for an 
in-plane current using the Gaussian GL action and the Kubo formula. 
As is well known, this is nothing but the $\rm Aslamasov-Larkin$ (AL) fluctuation 
conductivity\cite{AL}. 
After the analytical continuation, it is expressed by 

\begin{eqnarray} \label{eq:sigma}
	\sigma_{\rm s \perp} 
&=&
	\frac{2 \pi \xi_0^4}{R_QT}
	\int_{\om} 
	\frac{1}{\sinh^2(\frac{\omega}{2T})} 
	\int_{-\infty}^{\infty}
	\frac{{\rm d}^2 \q_\perp}{(2 \pi)^2} q_\perp^2 
	\int_{-\pi/d}^{\pi/d} \frac{{\rm d} q_z}{2\pi}
	\left[  {\rm Im} K^R(\q, \omega) \right]^2 \nonumber \\
&=&
	\frac{16 \pi \gamma_1 T}{3 R_Q \xi_{c0} \sqrt{\mu_0}} 
	\int_{\om}
	f \left( \mu_0 \omega/(2 \gamma_1 T) \right)
	 \int \frac{{\rm d}^3 \q}{(2 \pi)^3}  \frac{ q^2}
	{\left[ (1+q^2)^2 + \om^2 \right]^2}, 
\end{eqnarray}
where $R_Q=\pi \hbar/2e^2=6.45$ (k$\Omega$), 
$K^{R} (\q, \om ) = [{\cal K}^{-1}(q,0) - {\rm i} \gamma_1 \om ]^{-1}$ 
is the retarded fluctuation propagator, and, in moving to the last line, a 3D-type
behavior of fluctuation was assumed. 
This assumption is valid when ($0 <$) $\sqrt{\mu_0} < 2\xi_{c0}/s$. 
The weight of quantum fluctuation in 
$\sigma_{\rm s}$-expressions is measured by the function $f(x)=x^2/\sinh^2(x)$. 

First, let us comment on the strong disorder side $D > D_{\rm c}$ in which $\mu_0$ 
remains positive in $T \to 0$ limit. 
It is easily seen that $\sigma_{\rm s\perp}$ 
vanishes like\cite{Phil} $T^2$ in $T \to 0$ limit in 
this case. This is the simplest example of 
the {\it insulating}\cite{rf:Ikeda96} 
fluctuation conductivity at $T=0$. 
On the other hand, on the weak disorder side in which 
$D < D_{\rm c}$, $\mu_0$ vanishes at a positive temperature $T_{\rm c}(D)$ on cooling. 
Near $T_{\rm c}(D)$ where $\mu_0 \ll 2\gamma_1 T_{\rm c}(D)$, 
the familiar divergent (superconducting) behavior\cite{AL} 
$\sigma_{\rm s\perp}  \sim (T-T_{\rm c}(D))^{-1/2}$ is found. 
Now, let us consider $\sigma_{\rm s\perp}$ on 
the line $D=D_{\rm c}$ where $\mu_0 =  2\pi^2 (T\tau)^2/3$. 
The ratio $\mu_0/2\gamma_1T$ in this case also tends to vanish on cooling 
and becomes less than unity in $T < T_{\rm cr}^{(1)} \simeq 0.2 T_{\rm c0}$, 
and thus the same classical 
limit ($f(x) \to 1$) as in $D < D_{\rm c}$ dominates at $D=D_{\rm c}$ in low $T$ limit. 
Through a detailed analysis, 
we find the remaining quantum correction in $\sigma_{\rm s\perp}$ 
to be smaller by the factor $\sim \sqrt{T/T_{\rm c0}}$. 
Further, the condition of the 3D approximation, $\sqrt{\mu_0} < 2 \xi_{\rm c0}/s$, 
is rewritten in this case as $T < T_{\rm cr}^{(2)} \simeq \sqrt{3} t/(2\pi)$. 
Then, in $T \ll$ min($T_{\rm cr}^{(1)}$, $T_{\rm cr}^{(2)}$), 
$\sigma_{\rm s \perp}(D=D_{\rm c})$ approaches a $T$-independent value 

\begin{eqnarray} \label{eq:sigma_s}
	\sigma_{\rm s \perp}(D=D_{\rm c}) 
&=&
	\frac{1}{4 \pi}\sqrt{ \frac{3}{2}} \frac{ R_Q^{-1}}{\xi_{\rm c0}(D=D_{\rm c})} 
	\simeq \frac{\sqrt{3}}{7 R_Q \, s} \frac{T_{\rm c0}}{t}
\equiv
	\sigma_{\rm s}^*  
\end{eqnarray}
on cooling. By combining this with the $D \neq D_{\rm c}$ 
results mentioned above, one will notice that the situation around $D=D_{\rm c}$ 
in the present 3D case is similar to that of the 2D insulator-superconductor 
transition\cite{Phil}. A similar result is obtained for the out-of-plane 
conductivity $\sigma_{\rm s\parallel}$ of which the value at $D=D_{\rm c}$ is given by 

\begin{eqnarray}
	\sigma_{\rm s \parallel}(D=D_{\rm c}) 
&=&
	\sigma^*_{\rm s} \, \frac{\xi_{\rm c0}}{\xi_0} \biggr|_{D=D_{\rm c}} 
= 
	\frac{ts}{v_{\rm F}} \sigma_{\rm s\perp}(D=D_{\rm c}). 
\end{eqnarray}

Now, we consider the cases of pairing of $p$-wave type\cite{com}. 
Let us assume the attractive interaction to have a general form 
$V_{\rm BCS} ({\p}, {\p}') = |g|\sum_{j=x,y} w_j({\p}) w_j^*({\p}')$, 
where $w_j = p_j h_j(p_zs)$ with $ \int_{p_z} h_j(p_zs) > 0$. 
Then, the $d$-vector has the form 
$d_{\mu}(\p) = \sum_{j=x,y} p_j h_j(p_zs) \Delta_{\mu,j}$. 
This includes the candidates for Sr$_2$RuO$_4$ 
(triplet) pairing state,\cite{rf:Hasegawa00,Rice} 
$d_\mu(\p) = \sum_{\rho=\pm1} \Delta_\rho (p_x +{\rm i} \rho p_y) (c -{\rm cos}p_zs)$. 
As in this example, we will assume $|h_j(p_zs)|$ to be independent of $j$ 
for convenience of presentation. 
Under these apparatus, it is elementary to obtain the corresponding 
expressions to eqs. (\ref{eq:pi}) and (\ref{eq:vertex}) again in 
terms of the isotropic dispersion. The 
resulting Gaussian action is merely given here:
\begin{eqnarray} \label{eq:action_p}
	{\cal S}^{(p)} 
&=&
	\frac{N_2(0)}{2} \sum_{\mu, \atop i,j=x,y} \sum_{q,\om_{\nu}} 
	\Delta_{\mu ,i}^*(\q,\om_{\nu}) \, 
	[ 
	(\delta_{ij} - \hat{q}_i \hat{q}_j ) \, {\cal K}_{T}^{-1}(\q,\om_{\nu}) 
+ 	\hat{q}_i \hat{q}_j \, {\cal K}_{L}^{-1}(\q,\om_{\nu}) 
	] \, 
	\Delta_{\mu, j}(\q,\om_{\nu}), 	
\end{eqnarray}
where 
\begin{eqnarray}
	{\cal K}_{T}^{-1}(\q,\om_{\nu}) 
&=&
	\mu_0 + \gamma_1|\om_{\nu}| + \xi_0^2 q_\perp^2
	+ 4\left(\frac{\xi_{\rm c0}}{s}\right)^2 \sin^2 (q_z s/2), \nonumber \\
	{\cal K}_{L}^{-1}(\q,\om_{\nu})
&=&
	{\cal K}_{T}^{-1}(\q, \om_{\nu})
	+ \xi_0^2 C(q^2_\perp, T) \, q^2_\perp, 
\end{eqnarray}
${\hat q}_j = q_j/|\q_\perp|$, 
and 
$C(q^2_\perp, T)=2(1+\ln(4 \pi \tau T + q^2_\perp l^2)^{-1})$. 
The enhancement 
factor $C(q^2_\perp, T)$ arising from the diffusion propagator appears only in 
the longitudinal part of the lowest order terms in the gradient because the term 
corresponding to the second term of eq. (\ref{eq:vertex}) is vectorial and 
proportional to ${\q}$. An additional gradient term which may appear 
depending on the form of $\varepsilon_{\mibs p_\perp}$ will be discussed 
later. The in-plane conductivity $\sigma_{s \perp}$ 
in this case becomes

\begin{eqnarray} \label{eq:sigma_s^p}
	\sigma_{s \perp}^{(p)}(D=D_{\rm c}) 
&=&
	3 \times \frac{2 \pi \xi_0^4}{R_Q T} 
	\int \frac{{\rm d}^3 \q}{(2 \pi)^3} 
	\int_{\om} \frac{q_{\perp}^2}{ \sinh^2 \left(\frac{\om}{2T}\right) }
	\Bigg( \left( {\rm Im} K^R_{T}(\q,\om) \right)^2 +	(1+C(q_\perp^2, T))^2 
	\left( {\rm Im} K^R_{L}(\q,\om) \right)^2 \nonumber \\
&&	+ \frac{(C(q_\perp^2, T))^2}{2} 	
	\left( {\rm Im} K^R_{T}(\q,\om) \right) 	
	\left( {\rm Im} K^R_{L}(\q,\om) \right) 
	\Bigg)
	\biggr|_{D=D_{\rm c}} \nonumber \\
&\simeq&
	6 \sigma_{\rm s}^* 
	\left( 	\ln C(T) + {\rm const.} \right) 
\end{eqnarray}
on the line $D=D_{\rm c}$ and in low $T$ limit, 
where $C(T) = C(q^2_\perp=0, T) \simeq \ln(\tau T)^{-1}$, 
and the factor $3$ in the first line is due to spin degeneracy 
in the spin-triplet channel. 
In eq. (\ref{eq:sigma_s^p}), we have neglected contributions 
associated with an additional term of supercurrent arising from the 
$\q_\perp$-dependence of $C(q_\perp^2, T)$, 
which merely become, at most, of the order $\sigma_{\rm s}^* C^{-1}(T) \ln C(T)$. 
On the other hand, the out-of-plane conductivity is given by 
\begin{equation}
	\sigma_{s \parallel}^{(p)}(D=D_{\rm c}) \simeq 3 \sigma_{\rm s}^* 
	\left( 1 + C^{-1}(T) \right) \frac{\xi_{c0}}{\xi_0} \biggr|_{D=D_{\rm c}}.
\end{equation}

The low $T$ behaviors of $\sigma_{\rm s}^{(p)}$'s in $D \neq D_{\rm c}$ are 
similar to 
those in the case of non $p$-wave type: $\sigma_{\rm s}^{(p)}(D > D_{\rm c}) \to T^2$, 
while $\sigma_{\rm s}^{(p)}(D < D_{\rm c}) \sim (T-T_{\rm c}(D))^{-1/2}$ 
when $T \to T_{\rm c}(D)^+$. 
As eq. (\ref{eq:sigma_s^p}) shows, however, $\sigma_{s \perp}^{(p)}$ 
on the line $D=D_{\rm c}$ is, 
in contrast to eq. (\ref{eq:sigma_s}), weakly divergent 
(superconducting) in $T \to 0$ limit. 

We should note that it is the sum of a fluctuation contribution $\sigma_{\rm s}$ 
and a normal one $\sigma_{\rm n}$ which is directly measured in experiments 
of the type performed in ref. (\ref{eq:sigma_s^p}). 
If the electronic structure is of 3D-type, 
the residual normal conductivity is of the order 
$R_Q^{-1} E_{\rm F} \tau k_{\rm F}$, implying that the ratio 
$\sigma_{\rm s}^*/\sigma_{\rm n}$ will be small, O($(T_{\rm c0}/E_{\rm F})^2$), 
for 3D-type Fermi surfaces. 
In the present case with a quasi 2D Fermi surface, however, 
this ratio is larger: since 
$\sigma_{\rm n} \simeq R_Q^{-1} E_{\rm F} \tau/2s$, we have 
$\sigma^*_s/\sigma_{\rm n} \simeq \sqrt{6} s/(8 \pi E_{\rm F} 
\tau \xi_{\rm c0}(D=D_{\rm c}))$, 
which becomes $0.6 \pi \, T_{\rm c0}^2/(E_{\rm F} t)$ in the present model. 
In a system with $\xi_{\rm c0} \sim s$ (i.e., $t \sim T_{\rm c0}$) 
such as Sr$_2$RuO$_4$, 
this ratio is O($T_{\rm c0}/E_{\rm F}$). 
Although the ratio $T_{\rm c0}/E_{\rm F}$ in Sr$_2$RuO$_4$ is small ($\sim 10^{-2}$), 
it will be possible to separate $\sigma_{\rm s}$ from $\sigma_{\rm n}$ by applying 
a magnetic field as long as $\sigma_{\rm n}$ is insensitive to the field, 
because the fluctuation contribution $\sigma_{\rm s}$, 
in particular near $D=D_{\rm c}$, will be easily erased by applying a low magnetic 
field. 

A few comments on the above derivation of $\sigma_{\rm s}$ are in order. 
First, two main effects of crystalline anisotropy were neglected above. 
As one of them, the $\p$-dependence of the gap function $d_\mu(\p)$ was 
linearized above, and, as an example, $w_j(\p)$ should be proportional not to 
$p_j$ but to $\sin p_j$ in a real system. However, it is 
clear that the gradient terms, in particular its quadratic terms, of GL action 
do not essentially change due to this replacement. 
Next, the other gradient term which may appear in eq. (\ref{eq:action_p}) 
as a result of reduction of space 
symmetry should also be examined, which is typically given by 
\begin{equation} \label{eq:grad_invariant}
	|\partial_x \Delta_{\mu, x}|^2 + |\partial_y \Delta_{\mu, y}|^2 
\end{equation}
in zero field. The resulting GL gradient terms have three independent 
coefficients and hence, are essentially the same as those in the so-called 2D 
representation\cite{Vol} as long as the gauge field is 
spatially constant (note that a conductance for a uniform current in zero 
magnetic field is defined in terms of a spatially constant gauge field). 
However, the inclusion of eq. (\ref{eq:grad_invariant}) does not qualitatively 
change our main results (\ref{eq:sigma_s}) and (\ref{eq:sigma_s^p}). 
A key point for reaching this conclusion is to notice that the 
contribution accompanied by the diffusion propagator leads not to the term 
(15) but only to a longitudinal term $\sim |{\rm div}{\bf \Delta}|^2$ and that, 
as in clean limit, the term (\ref{eq:grad_invariant}) can arise simply as a combined 
result of crystal anisotropy 
in the dispersion $\varepsilon_{\mibs p_\perp}$ and of 
the $w_j({\p})$-form. 
Then, it is straightforward to see, by power 
counting in $\q$-integrals, the appearance in cases of $p$-wave type of a 
$\ln \ln(T\tau)^{-1}$-divergence like that in eq. (\ref{eq:sigma_s^p}) 
even when the term (\ref{eq:grad_invariant}) is present. 

We have focused above on the AL term of Gaussian fluctuation conductivity. 
It is not difficult to examine other contributions, $\rm Maki-Thompson$\cite{Tink} 
term $\delta \sigma_{\rm MT}$ and the so-called DOS term $\delta \sigma_{\rm DOS}$. 
We find that, in $T \to 0$ limit, they vanish on the $D=D_{\rm c}$ line 
in the manners $\delta \sigma_{\rm MT} \sim T\tau$ and 
$\delta \sigma_{\rm DOS} \sim (T\tau)^2$, and thus, 
we can neglect them as far as the behavior in low $T$ limit is concerned. 

Fluctuation conductivity near the disorder-induced quantum critical point
 was also examined in ref.15, where the $p$-wave case does not seem to have been 
considered. 
The corresponding result to eq. (\ref{eq:sigma_s}) was concluded in ref. 15 
to be rather 
proportional to $T^{1/4}$. This difference originates from the assumption 
implicit in ref. 15 that the mass renormalization $\delta \mu \sim T^{D/2}$ 
in $D$-dimensions in the one-loop order\cite{Millis} is larger than the bare 
one $\mu_0 \sim T^2$. However, the former arises from the quartic 
(nonGaussian) term of GL action. In fact, it is 
verified that the above assumption $\delta\mu > \mu_0$ is valid 
in $T < T_{\rm c0} \varepsilon_{\rm G}^{(3)}$, where $\varepsilon_{\rm G}^{(3)} 
\simeq (T_{\rm c0} s/(E_{\rm F} \xi_{\rm c0}))^2$ is the 3D 
Ginzburg number measuring the temperature width $\Delta T/T_{\rm c0}$ 
of the {\it thermal critical} region. Namely, if this temperature width 
of O($T_{\rm c0}^3/E_{\rm F}^2$) is negligible as in Sr$_2$RuO$_4$, 
the $T^{1/4}$-behavior of $\sigma_{\rm s}(D=D_{\rm c})$ will not be observable at 
accessible temperatures, 
but rather, the present Gaussian results should be observed at low $T$. 

We emphasize that the result, eq. (\ref{eq:sigma_s^p}), 
in cases of $p$-wave type is valid only when the Fermi surface and thus, 
the gap function $d({\p})$ are of quasi 2D-type. 
If the Fermi surface and hence, $d({\p})$ is essentially 3D-type 
({\it e.g.}, $d({\p})=\sum_{j=x,y,z} \Delta_j \, p_j$), 
$\sigma_{\rm s}(D=D_{\rm c}; \, T \to 0)$ 
is nondivergent and is not useful in distinguishing the pairing states. 
By contrast, if $d({\p})$ is 1D-type and, as its example, 
is approximated by ${\rm sin}(p_x)$, the 
fluctuation conductivity $\sigma_{\rm s \, x}(D=D_{\rm c})$ in the $x$-direction is 
divergent like $(\ln(1/T\tau))^{1/2}$, while the corresponding ones in the $y$ 
and $z$ directions instead vanish like $(\ln(1/T\tau))^{-1/2}$ in low $T$ 
limit. 

In conclusion, the fluctuation conductivities near a 
disorder-induced quantum critical point of a non $s$-wave bulk superconductor 
with quasi 2D-like electronic structure were examined in the Gaussian 
approximation 
and in zero field. When the dependence of the gap function on 
the {\it in-plane} momentum $\p_\perp$ is of a higher order, as in cases of 
$d$-wave or $f$-wave symmetry, $\sigma_{\rm s}$ at the crtitical 
value of disorder approaches a finite value in low $T$ limit, just like the 2D 
nonGaussian result,\cite{Phil} while it 
is weakly divergent on cooling when the 
gap function is linear in $\p_\perp$ or $\sin (\p_{\perp})$ 
(of a $p$-wave type). 
The present result may be useful for clarifying the form of pairing function 
of a material with spin-triplet pairing and a low dimensional Fermi surface 
and, for example, clarifying which of the $f$-wave gap function\cite{rf:Hasegawa00} 
$\propto p_x \pm {\rm i} p_y$ and the function\cite{Balatsky} $\propto 
p_x p_y (p_x \pm {\rm i} p_y)$ is favorable as the pairing 
state of Sr$_2$RuO$_4$. 

We are grateful to Y. Maeno and an anonymous referee for comments. This work 
was supported by CREST of Japan Science and Technology Corporation (JST).

\vspace{50mm}


\end{document}